\documentclass[conference]{IEEEtran}
\IEEEoverridecommandlockouts

\usepackage{cite}
\usepackage{amsmath,amssymb,amsfonts}
\usepackage{algorithmic}
\usepackage{graphicx}
\usepackage{textcomp}
\usepackage{xcolor}
\usepackage{booktabs}
\usepackage{subcaption}
\usepackage{stfloats}
\usepackage{hyperref}
\usepackage[linesnumbered,ruled,vlined]{algorithm2e}
\usepackage{balance}

\def\BibTeX{{\rm B\kern-.05em{\sc i\kern-.025em b}\kern-.08em
		T\kern-.1667em\lower.7ex\hbox{E}\kern-.125emX}}
\renewcommand{\baselinestretch}{0.972}  %
\begin{document}

	\title{Knowledge Distillation Driven Semantic NOMA with GAN Refinement for 6G Robotic Vehicle Networks\\
	}

	\author{
		\IEEEauthorblockN{
			Qifei Wang\textsuperscript{1},
			Zhen Gao\textsuperscript{1,2,*}, %
			Li Qiao\textsuperscript{3},      %
			Ziwei Wan\textsuperscript{2},
			De Mi\textsuperscript{4},        %
			Dapeng Li\textsuperscript{1}
			and Ying Sun\textsuperscript{1}
		}
		\IEEEauthorblockA{\textsuperscript{1}School of Information and Electronics, Beijing Institute of Technology, Beijing 100081, China}
		\IEEEauthorblockA{\textsuperscript{2}The Yangtze Delta Region Academy, Beijing Institute of Technology, Jiaxing 314019, China}
		\IEEEauthorblockA{\textsuperscript{3}Department of Electrical and Electronic Engineering, The University of Hong Kong, Pokfulam Road, Hong Kong}
		\IEEEauthorblockA{\textsuperscript{4}College of Computing, Birmingham City University, Birmingham, B4 7XG, U.K}
		\IEEEauthorblockA{*Corresponding author}
		\IEEEauthorblockA{Email: \{qfwang, gaozhen16, ziweiwan, dapangli, sunyingsy\}@bit.edu.cn, qiaoli@hku.hk, De.Mi@bcu.ac.uk}

	}
	\IEEEaftertitletext{\vspace{-1.4\baselineskip}} %
	\maketitle
	\begin{abstract}
		To achieve sustainable intelligent mobility, 6G-empowered robotic vehicles (RVs) require high-fidelity visual perception under stringent bandwidth and energy constraints. Semantic communication offers a spectral-efficient solution but suffers from severe interference in uplink non-orthogonal multiple access (NOMA) RV networks. To address this, we propose a knowledge distillation-driven and generative models-enhanced NOMA framework for robust and green RV communications, named KDG-SemNOMA. First, we develop a ConvNeXt-based deep joint source-channel coding (DeepJSCC) architecture with an enhanced attention feature (AF) module for dynamic channel adaptation. Second, to mitigate interference without inference overhead, an orthogonal transmission teacher model guides the NOMA student model via a two-stage knowledge distillation strategy. Finally, to address the over-smoothing artifacts of pixel-wise optimization, we introduce a channel-conditional GAN (cGAN). By explicitly taking the Stage-I initial reconstruction and channel states as conditional inputs, this module refines coarse outputs into high-fidelity images with realistic textures. Experiments on FFHQ-256 demonstrate that KDG-SemNOMA significantly outperforms state-of-the-art methods in both pixel-level accuracy and perceptual fidelity.
	\end{abstract}

	\begin{IEEEkeywords}
		6G robotic vehicles, semantic communication, NOMA, knowledge distillation, conditional GAN.
	\end{IEEEkeywords}
	\vspace{-3mm}
	\section{Introduction}
	Robotic vehicles (RVs) are envisioned as a key enabler for sustainable 6G mobility systems. However, the massive visual data generated by RVs for remote control and situational awareness imposes a heavy burden on bandwidth and energy resources \cite{dhyvehicle}. To achieve eco-friendly and reliable connectivity, semantic communication \cite{bourtsoulatze2019deep} has emerged as a paradigm shift, transmitting semantic features rather than raw bits to significantly improve transmission efficiency.

	Recent research has actively expanded semantic communication into multi-user and generative domains. On the one hand, several works have explored multi-user semantic transmission. For instance, the work \cite{yilmaz2023distributed} proposed a pioneering deep joint source-channel coding (DeepJSCC) method for uplik non-orthogonal multiple access (NOMA) transmission, named DeepJSCC-NOMA. The paper \cite{zhang2024interference} have investigated interference cancellation strategies to manage multi-user semantic features. The work \cite{yanadaptive} propose semantic adaptive feature extraction network for downlink NOMA transmission. The work \cite{mengprompt} integrates generative adversarial networks (GAN)-based interference mitigation. On the other hand, to improve reconstruction quality beyond simple distortion metrics, GAN models have been integrated into DeepJSCC systems. These approaches \cite{erdemir2023generative} leverage generative priors to enhance perceptual fidelity, compensating perceptual information loss.

	Despite these advances, deploying DeepJSCC in 6G RV networks faces three critical challenges. First, existing multi-user semantic systems exhibit insufficient interference cancellation capabilities for non-orthogonal signals. Second, current schemes fail to adapt to the varying channel conditions in RVs, neglecting channel state information (CSI) including both signal-to-noise ratio (SNR) fluctuations and specific fading parameters (e.g., amplitude and phase). Third, pixel-wise optimization inherently results in blurry reconstructions, which hinders precise visual-based navigation and decision-making.

	To address these dual challenges of interference mitigation and perceptual quality enhancement, we propose a novel framework termed KDG-SemNOMA\footnote{This conference paper has been expanded and submitted to possible journal \cite{wang2025knowledge}.}. The main contributions of this paper are summarized as follows:
	\begin{itemize}
		\item \textbf{Robust SemNOMA Architecture:} We design a multi-user semantic transceiver based on the ConvNeXt backbone. Crucially, we introduce an enhanced attention feature (AF) module that explicitly leverages CSI, including SNR and Rayleigh fading parameters, to dynamically modulate semantic features, thereby enhancing robustness against dynamic channel variations.

		\item \textbf{Knowledge Distillation (KD) Optimization:} To effectively mitigate multi-user interference, we propose a two-stage KD strategy. We utilize an "interference-free" model trained under orthogonal multiple access (OMA) as the \textit{Teacher}, guiding the NOMA-based \textit{Student} model. By employing feature affinity (FA) and cross-head distillation (CrossKD) losses, the student learns to extract clean semantic features from superimposed signals without incurring additional inference overhead.

		\item \textbf{Channel-Conditional GAN Refinement:} To overcome the blurriness inherent in mean absolute error (MAE)-based optimization, we introduce a second-stage refinement module based on a conditional GAN (cGAN). Specifically, the generator takes the concatenation of the Stage-I initial reconstruction and the spatial channel state map as input. This explicit conditioning enables the model to adaptively hallucinate high-frequency textures and restore perceptual fidelity commensurate with the specific channel degradation intensity.
	\end{itemize}

	The remainder of this paper is organized as follows. Section \ref{System_model}introduces the system model and the proposed network architecture. Section \ref{sec:kd} details the knowledge distillation strategy for interference mitigation. Section \ref{sec:cgan} presents the channel-conditional GAN for image refinement. Section \ref{simulation} provides the simulation results and performance analysis, followed by the conclusion in Section \ref{conclusion}.
	\vspace{-2.3mm}
	\section{System Model and Network Architecture} \label{System_model}
	\begin{figure*}[!t]
		\centering
		\includegraphics[width=0.90\linewidth]{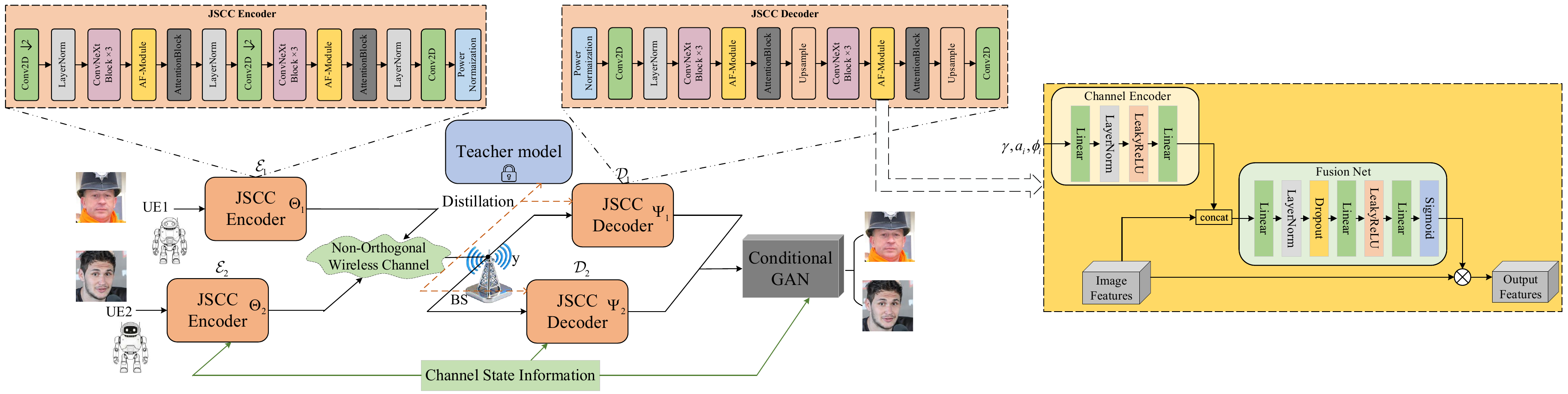}
		\captionsetup{font={footnotesize},justification=raggedright,singlelinecheck=false}
		\caption{The system model for our KDG-SemNOMA framework.}
		\label{Fig1}
		\vspace{-7mm}
	\end{figure*}
	\subsection{Semantic NOMA Transmission Model}
	As illustrated in Figure \ref{Fig1}, we consider an uplink multi-user semantic communication system where $N$ robotic vehicle user equipment (UE) devices transmit images to a single base station (BS). Let $\mathbf{x}_i \in \mathbb{R}^{C \times H \times W}$ denote the source image of the $i$-th UE, where $C, H, W$ represent the number of channels, height, and width, respectively.

	To enable the BS to distinguish between users in the NOMA superposition, we adopt a user-specific embedding strategy. A unique identification vector $\mathbf{r}_i \in \mathbb{R}^{1 \times H \times W}$ is concatenated with the input image along the channel dimension. The combined input is fed into the semantic encoder $\mathcal{E}_i(\cdot; \boldsymbol{\Theta}_i)$:
	\begin{equation}
		\mathbf{s}_i = \mathcal{E}_i(\text{Concat}(\mathbf{x}_i, \mathbf{r}_i); \boldsymbol{\Theta}_i),
		\vspace{-1mm}
	\end{equation}
	where $\mathbf{s}_i \in \mathbb{C}^k$ is the complex-valued semantic symbol sequence, and $k$ is the number of channel uses. The bandwidth compression ratio is defined as $\rho = k/m$.

	To satisfy the power constraints, the transmitted symbols are normalized to satisfy the average power constraint $P_{avg}$:
	\begin{equation}
		\frac{1}{k} \mathbb{E}[\|\mathbf{s}_i\|_2^2] \le P_{avg}, \quad \forall i.
		\vspace{-1mm}
	\end{equation}

	In the uplink NOMA scenario, users transmit simultaneously over the same time-frequency resources. The received signal $\mathbf{y} \in \mathbb{C}^k$ at the BS is the superposition of signals from all users distorted by the wireless channel:
	\begin{equation}
		\mathbf{y} = \sum_{i=1}^{N} h_i \mathbf{s}_i + \mathbf{n},
		\label{eq:noma_transmission}
		\vspace{-2mm}
	\end{equation}
	where $h_i \in \mathbb{C}$ denotes the channel gain coefficient for the $i$-th UE, and $\mathbf{n} \sim \mathcal{CN}(0, \sigma^2 \mathbf{I}_k)$ represents the additive white Gaussian noise (AWGN) with noise power $\sigma^2$. We consider both AWGN channels ($h_i=1$) and Rayleigh fading channels, where $h_i \sim \mathcal{CN}(0, 1)$. The SNR is defined as $\gamma = 10 \log_{10}(P_{avg}/\sigma^2)$.

	Finally, the BS employs semantic decoder $\mathcal{D}_i(\cdot; \boldsymbol{\Psi})$ to reconstruct the source images from the received signal $\mathbf{y}$:
	\begin{equation}
		\hat{\mathbf{x}}_i = \mathcal{D}_i(\mathbf{y};\boldsymbol{\Psi}_i).
		\vspace{-2mm}
	\end{equation}

	\subsection{ConvNeXt-based Architecture with Enhanced AF-Module}
	The transceiver backbone adopts the ConvNeXt \cite{liu2022convnet} architecture for its superior feature extraction capabilities. Inspired by \cite{xu2021wireless}, to ensure robustness against dynamic fading, we integrate an Enhanced Attention Feature (AF) Module (Figure \ref{Fig1}).
	Unlike standard attention mechanisms, the AF-Module explicitly conditions feature modulation on the instantaneous channel state. We define the channel statistics vector as $\mathbf{v}_{csi} = [\gamma, a_i, \phi_i]$, encapsulating SNR, fading amplitude, and phase, respectively.
	By projecting $\mathbf{v}_{csi}$ into a channel embedding and fusing it with global context features, the module generates a channel-wise attention mask $\mathbf{M}$. The refined features are obtained via:
	\begin{equation}
		\mathbf{F}_{out} = \mathbf{F}_{in} \odot \text{Broadcast}(\mathbf{M}).
		\vspace{-2mm}
	\end{equation}
	This mechanism allows the network to adaptively emphasize features that are robust to the specific current channel conditions while suppressing noise-sensitive components.
	\vspace{-2.5mm}
	\section{Knowledge Distillation Optimization for Semantic NOMA (KD-SemNOMA)}
	\label{sec:kd}
	\begin{figure*}[!t]
		\centering
		\includegraphics[width=0.80\linewidth]{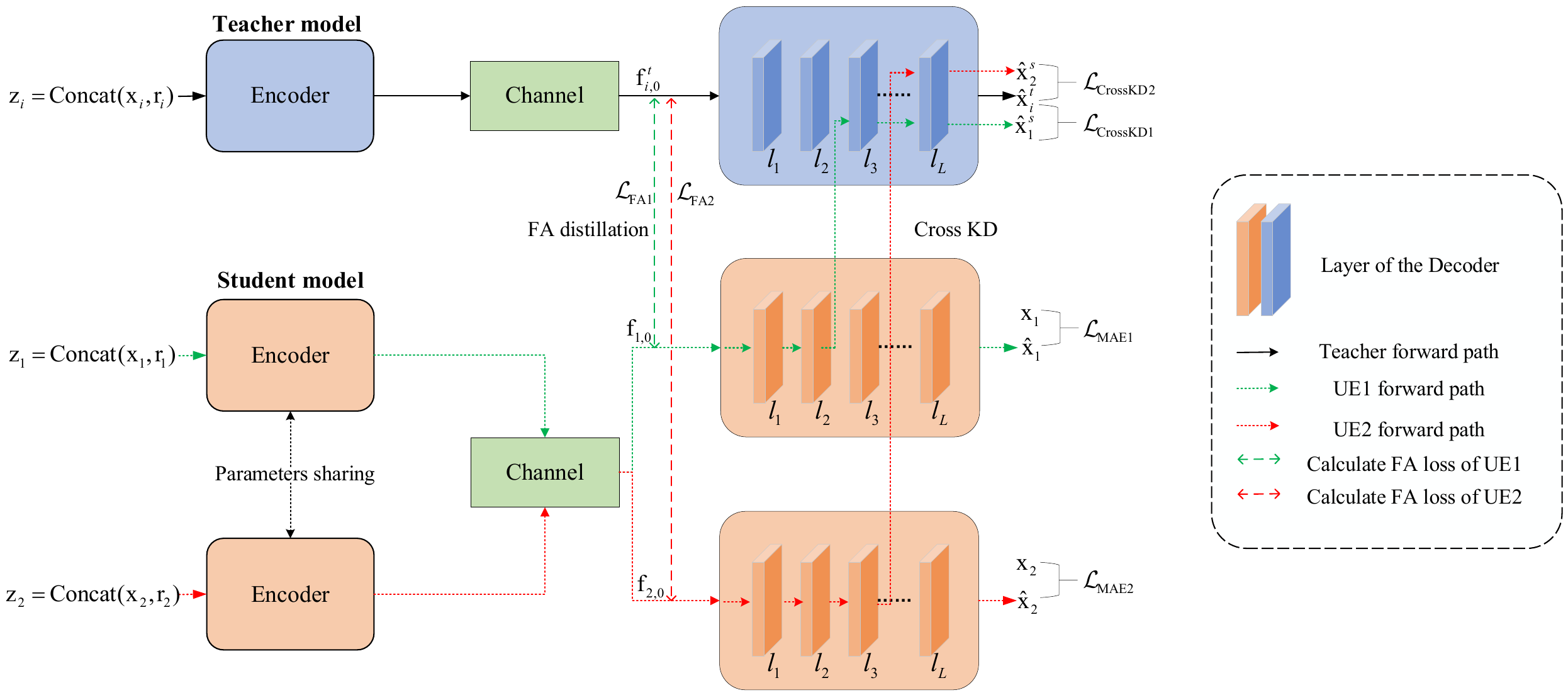}
		\captionsetup{font={footnotesize},justification=raggedright,singlelinecheck=false}
		\caption{The overall framework of KD-SemNOMA. $l_j$ denotes the $j$-th layer of the decoder. $\hat{x}_i$ denotes the student model output of the $i$-th UE, $\mathbf{f}_{i,0}^t$ denotes the input feature of the teacher decoder, $\mathcal{L}_{\text{MAEi}}$ denotes the MAE loss of the $i$-th UE, $\mathcal{L}_{\text{CrossKDi}}$ denotes the CrossKD loss of the $i$-th UE, $\mathcal{L}_{\text{FAi}}$ denotes the FA loss of the $i$-th UE.}
		\label{Fig3}
		\vspace{-5mm}
	\end{figure*}
	To mitigate NOMA interference without increasing inference complexity, we propose KD-SemNOMA. As shown in Fig. \ref{Fig3}, an \emph{interference-free} OMA teacher guides a NOMA student. The teacher model, trained on clean orthogonal channels, provides robust semantic supervision for the student model, which operates under shared-resource interference. During inference, only the student model is deployed, which incurs no additional computational overhead.

	\subsection{Teacher Model Training}
	For the $i$-th UE, the teacher network transmits semantic features $\mathbf{s}_i$ over an interference-free orthogonal channel
	\begin{equation}
		\mathbf{y}_i = h_i \mathbf{s}_i + \mathbf{n},
		\label{eq:orthogonal_transmission_conf}
		\vspace{-2mm}
	\end{equation}
	The teacher model $\mathcal{F}_{\text{teacher}}(\cdot)$ is optimized with the MAE reconstruction loss \cite{zou2020deep}
	\begin{equation}
		\mathcal{L}_{T} = \sum_{i=1}^{N}
		\big\|\mathcal{F}_{\text{teacher}}(\mathbf{x}_i) - \mathbf{x}_i \big\|_1,
		\label{eq:teacher_loss_conf}
		\vspace{-2mm}
	\end{equation}
	where $\mathbf{x}_i$ denotes the input image of the $i$-th UE.
	\vspace{-4mm}
	\subsection{Student Model Training}

	The student SemNOMA model $\mathcal{F}_{\text{student}}(\cdot)$ is initialized from the pre-trained teacher. It is trained under the NOMA channel model in Section \ref{System_model} and supervised by three complementary losses.

	\subsubsection{Feature Affinity Distillation}

	Feature affinity (FA) distillation \cite{he2020fakd} transfers structural semantic information from the teacher model to student model. For selected decoder layers, we construct a spatial affinity matrix from the feature maps of teacher and student, and minimize their difference. Denoting the affinity matrices of the $i$-th user from the teacher and student as $\mathbf{A}_{i,j}^t$ and $\mathbf{A}_{i,j}^s$ at layer $j$, the FA loss is
	\begin{equation}
		\mathcal{L}_{\mathrm{FA}} =
		\sum_{i=1}^{N} \frac{1}{|\mathbf{A}_i|}
		\sum_{j} \big\| \mathbf{A}_{i,j}^{s} - \mathbf{A}_{i,j}^{t} \big\|_1,
		\label{eq:fa_loss_conf}
		\vspace{-2mm}
	\end{equation}
	where $|\mathbf{A}_i|$ denotes the number of elements in the affinity matrix of the $i$-th user.

	\subsubsection{Cross-Head Prediction Distillation}

	To further align the high-level semantics, we apply Cross-Head Knowledge Distillation (CrossKD) \cite{wang2024crosskd}. Specifically, the intermediate decoder features of the student are fed into the teacher’s decoder head to obtain a cross prediction $\hat{\mathbf{x}}_i^{\text{cross}}$, which is then enforced to match the clean reconstruction $\hat{\mathbf{x}}_i^{t}$ of the teacher:
	\begin{equation}
		\mathcal{L}_{\text{CrossKD}} =
		\sum_{i=1}^{N}
		\big\| \hat{\mathbf{x}}_i^{\text{cross}} - \hat{\mathbf{x}}_i^{t} \big\|_1.
		\label{eq:crosskd_loss_conf}
		\vspace{-2mm}
	\end{equation}
	This encourages the student features to be compatible with the teacher’s decision boundaries without directly forcing their outputs to coincide.

	\subsubsection{Reconstruction Loss and Total Objective}

	Combining the standard reconstruction loss $\mathcal{L}_{\text{MAE}}$, the overall objective for the student is a weighted sum of the three losses:
	\begin{equation}
		\mathcal{L}_{S} =
		\lambda_1 \mathcal{L}_{\text{MAE}} +
		\lambda_2 \mathcal{L}_{\mathrm{FA}} +
		\lambda_3 \mathcal{L}_{\text{CrossKD}},
		\label{eq:total_loss_conf}
		\vspace{-2mm}
	\end{equation}
	where $\lambda_1$, $\lambda_2$, and $\lambda_3$ balance reconstruction fidelity and distillation strength.

	\vspace{-2.5mm}
	\section{Stage-II: Conditional GAN-based Channel-Aware Refinement}
	\label{sec:cgan}

	\begin{figure}[!t]
		\centering
		\includegraphics[width=3.3in]{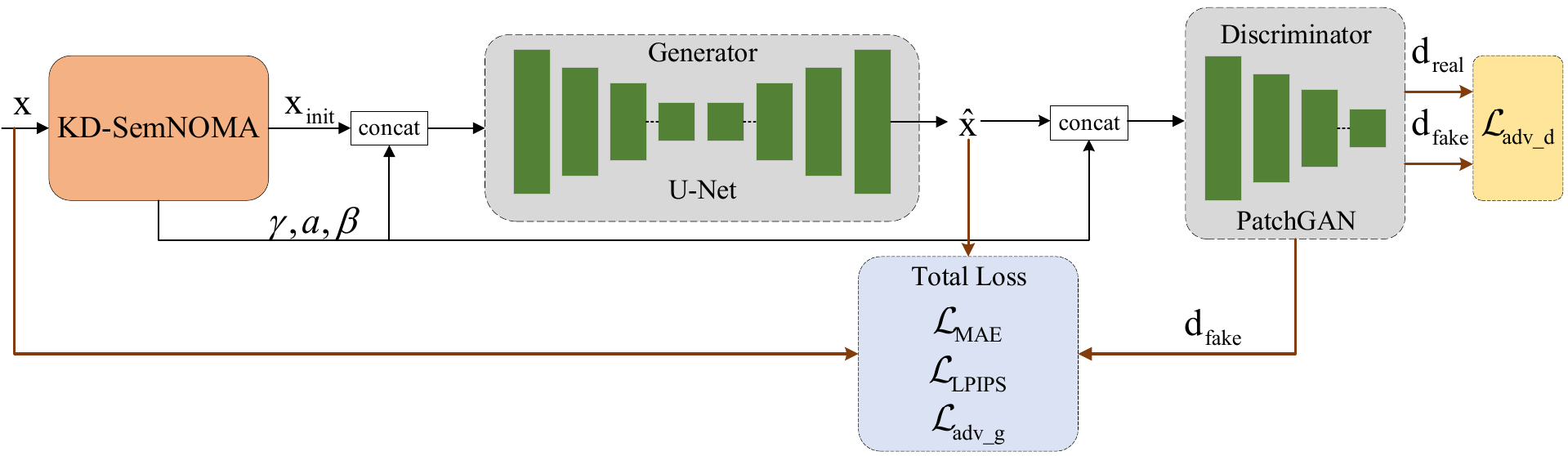}
		\captionsetup{font={footnotesize},justification=raggedright,singlelinecheck=false}
		\caption{Architecture of the proposed Stage-II image refinement framework utilizing a channel-conditional GAN.}
		\label{Fig:cgan_framework}
		\vspace{-7mm}
	\end{figure}

	Building upon the coarse reconstruction yielded by the Stage-I KD-SemNOMA model (denoted as $\mathcal{M}_S$), the objective of Stage-II is to restore high-frequency textural details. As illustrated in Figure \ref{Fig:cgan_framework}, crucially, this refinement is explicitly conditioned on the instantaneous channel quality to adaptively mitigate artifacts. %

	Let $\mathbf{x} \in \mathbb{R}^{3 \times H \times W}$ represent the ground-truth clean image. The Stage-I model generates a preliminary reconstruction $\hat{\mathbf{x}}_{init}$:
	\begin{equation}
		\hat{\mathbf{x}}_{init} = \mathcal{M}_S(\mathbf{x}, \mathbf{v}_{csi}),
		\vspace{-2mm}
	\end{equation}
	where $\hat{\mathbf{x}}_{init} \in \mathbb{R}^{3 \times H \times W}$ preserves the fundamental semantic layout but lacks fine-grained details. The vector $\mathbf{v}_{csi} \in \mathbb{R}^{d_s}$ encapsulates channel statistics, such as SNR and fading parameters. Specifically, $d_s=1$ is used for AWGN channels, while $d_s=3$ is employed for Rayleigh fading scenarios to capture amplitude and phase characteristics.

	To integrate the low-dimensional CSI vector into the convolutional architecture, we project $\mathbf{v}_{csi}$ via a linear layer $\phi: \mathbb{R}^{d_s} \rightarrow \mathbb{R}^{3}$ and subsequently expand it spatially. This results in a feature map $\mathbf{M}_{csi} \in \mathbb{R}^{3 \times H \times W}$ matching the image resolution:
	\begin{equation}
		\mathbf{M}_{csi} = \text{Broadcast}(\phi(\mathbf{v}_{csi})).
		\vspace{-2mm}
	\end{equation}

	\subsection{Conditional Generator and Discriminator}
	\vspace{-1mm}
	The refinement framework is implemented as a cGAN. The Generator, $G$, receives a concatenation of the initial estimate and the channel feature map to synthesize the refined image $\hat{\mathbf{x}}$:
	\begin{equation}
		\hat{\mathbf{x}} = G\big( \text{Concat}(\hat{\mathbf{x}}_{init}, \mathbf{M}_{csi}) \big).
		\vspace{-2mm}
	\end{equation}

	The Discriminator, $D$, functions as a critic that evaluates the authenticity of the input image conditioned on the channel state $\mathbf{M}_{csi}$. The critic outputs scores for real and generated pairs as follows:
	\begin{align}
		D_{\text{real}} &= D\big( \text{Concat}(\mathbf{x}, \mathbf{M}_{csi}) \big), \\
		D_{\text{fake}} &= D\big( \text{Concat}(\hat{\mathbf{x}}, \mathbf{M}_{csi}) \big).
		\vspace{-2mm}
	\end{align}
	By conditioning $D$ on $\mathbf{M}_{csi}$, the generator is compelled to learn adaptive restoration strategies specific to different channel distortions (e.g., distinguishing between heavy fading and mild noise).
	\vspace{-2.5mm}
	\subsection{Loss Functions}
	\subsubsection{Generator Objective}
	To guarantee both pixel-wise accuracy and perceptual quality, $G$ is optimized using a composite loss function:
	\begin{equation}
		\begin{aligned}
			\mathcal{L}_G &= \lambda_{\text{MAE}} \mathcal{L}_{\text{MAE}} + \lambda_{\text{LPIPS}} \mathcal{L}_{\text{LPIPS}} + \lambda_{\text{adv}} \mathcal{L}_{\text{adv}}^G,
		\end{aligned}
		\vspace{-1.5mm}
	\end{equation}
	where $\mathcal{L}_{\text{MAE}}$ enforces low-frequency consistency, $\mathcal{L}_{\text{LPIPS}}$ minimizes perceptual discrepancies in the VGG feature space, and $\mathcal{L}_{\text{adv}}^G = - \mathbb{E}[ D([\hat{\mathbf{x}}, \mathbf{M}_{csi}]) ]$ represents the Wasserstein adversarial loss.

	\subsubsection{Discriminator Objective}
	The discriminator aims to maximize the Wasserstein distance between the real and synthesized data distributions. This is achieved by minimizing the following objective:
	\begin{equation}
		\mathcal{L}_D = \mathbb{E}[D_{\text{fake}}] - \mathbb{E}[D_{\text{real}}].
		\vspace{-1.5mm}
	\end{equation}
	The optimization is performed via alternating gradient updates between $D$ and $G$.

	\vspace{-3.0mm}
	\section{Simulation Results} \label{simulation}
	\vspace{-1mm}
	\subsection{Simulation Setup}
	\subsubsection{Datasets and Implementation} We evaluate our framework on the high-resolution FFHQ dataset \cite{karras2019style}. For training and testing, the images are downsampled to $256 \times 256$ resolution, simulating multi-user transmission under AWGN and Rayleigh fading channels with SNR uniformly sampled from $[0, 20]$ dB. The framework is implemented in PyTorch on an NVIDIA 4090 GPU using the AdamW optimizer ($lr=10^{-4}$). For \textbf{Stage-I}, the KD-SemNOMA student model is trained with loss weights $\lambda_1=10, \lambda_2=100, \lambda_3=1$. For \textbf{Stage-II}, we freeze Stage-I and train the cGAN (U-Net generator, PatchGAN discriminator), with MAE, Learned Perceptual Image Patch Similarity (LPIPS), and adversarial loss weights set to $1, 0.1$, and $10^{-4}$, respectively.

	\subsubsection{Benchmark Schemes}
	We compare our method against three baselines: 1) \textbf{BPG+LDPC+QAM+SIC}: A conventional separation-based scheme employing BPG for source coding, LDPC (1/2 rate) for channel coding, 4QAM modulation, and SIC detection; 2) \textbf{DeepJSCC-NOMA}: An attention-based semantic baseline \cite{yilmaz2023distributed}; 3) \textbf{SemOMA}: An orthogonal transmission benchmark evaluated at both \textit{equivalent} and \textit{double} transmission overheads to serve as interference-free performance bounds.

	\subsubsection{Metrics}
	We adopt Peak Signal-to-Noise Ratio (PSNR) to evaluate pixel-level fidelity. To assess perceptual quality, we use LPIPS and Fréchet Inception Distance (FID).

	\vspace{-2mm}
	\subsection{Performance Analysis}
	\begin{figure}[!t]
		\centering
		\begin{subfigure}[t]{0.48\linewidth}
			\centering
			\includegraphics[width=\linewidth]{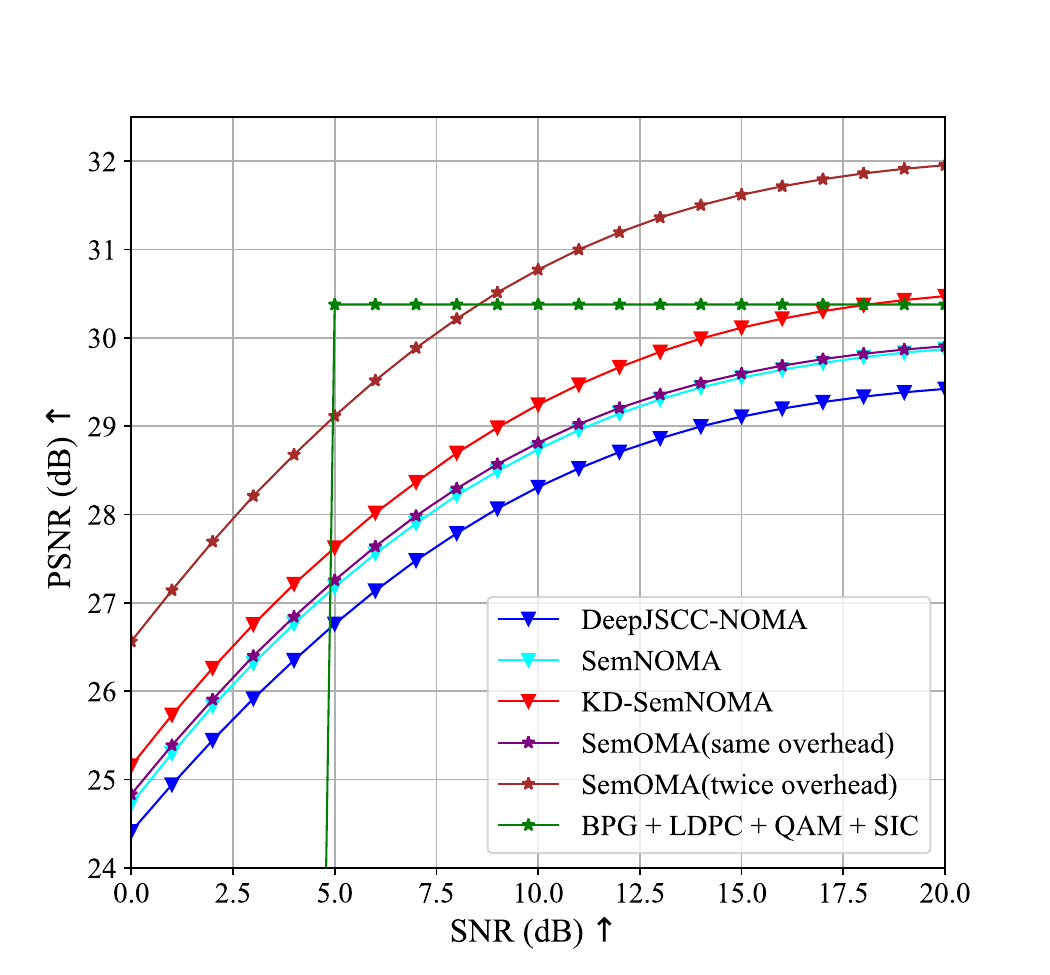}
			\captionsetup{font={footnotesize}}
			\caption{PSNR vs. SNR (AWGN)}
			\label{fig:psnr_awgn}
		\end{subfigure}
		\hfill %
		\begin{subfigure}[t]{0.48\linewidth}
			\centering
			\includegraphics[width=\linewidth]{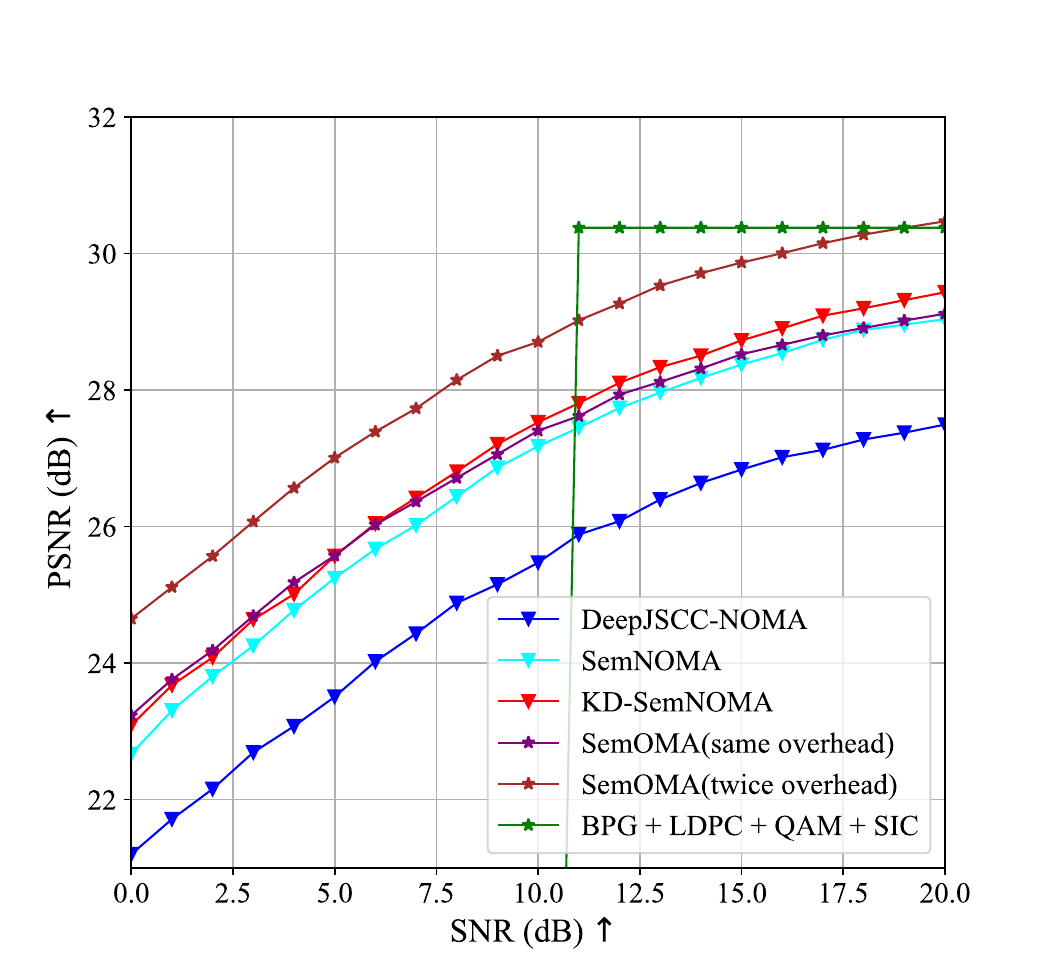}
			\captionsetup{font={footnotesize}}
			\caption{PSNR vs. SNR (Rayleigh)}
			\label{fig:psnr_rayleigh}
		\end{subfigure}

		\captionsetup{font={footnotesize,{color=black}}}
		\caption{PSNR performance comparison on FFHQ-256 dataset (2UE, $\rho=1/48$). Subfigure (a) shows the performance under AWGN channel, and (b) shows the performance under Rayleigh fading channel.}
		\label{fig:performance_psnr}
		\vspace{-7mm} %
	\end{figure}

	\begin{figure*}[!t]
		\centering
		\begin{subfigure}[t]{0.48\columnwidth}
			\centering
			\includegraphics[width=\linewidth]{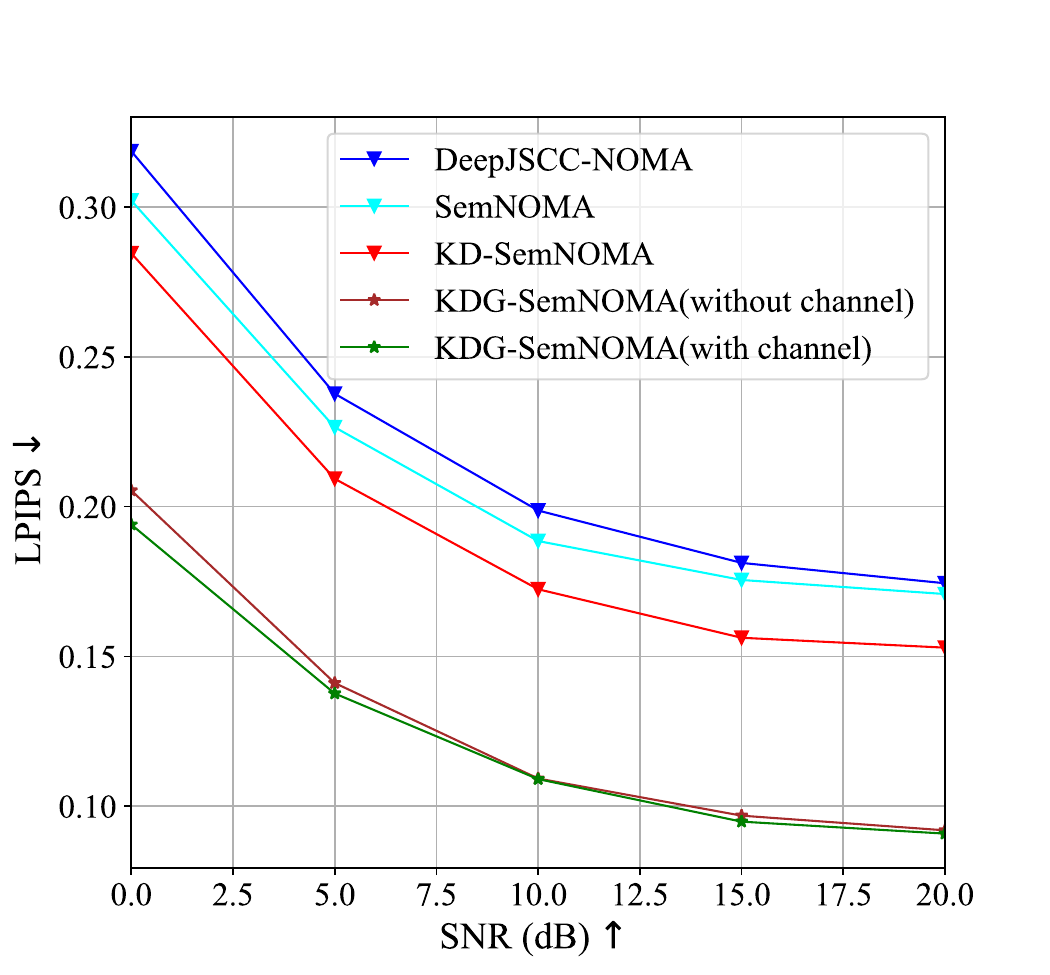}
			\captionsetup{font={footnotesize}}
			\caption{LPIPS vs. SNR (AWGN)}
			\label{fig8a}
		\end{subfigure}
		\hfill
		\begin{subfigure}[t]{0.48\columnwidth}
			\centering
			\includegraphics[width=\linewidth]{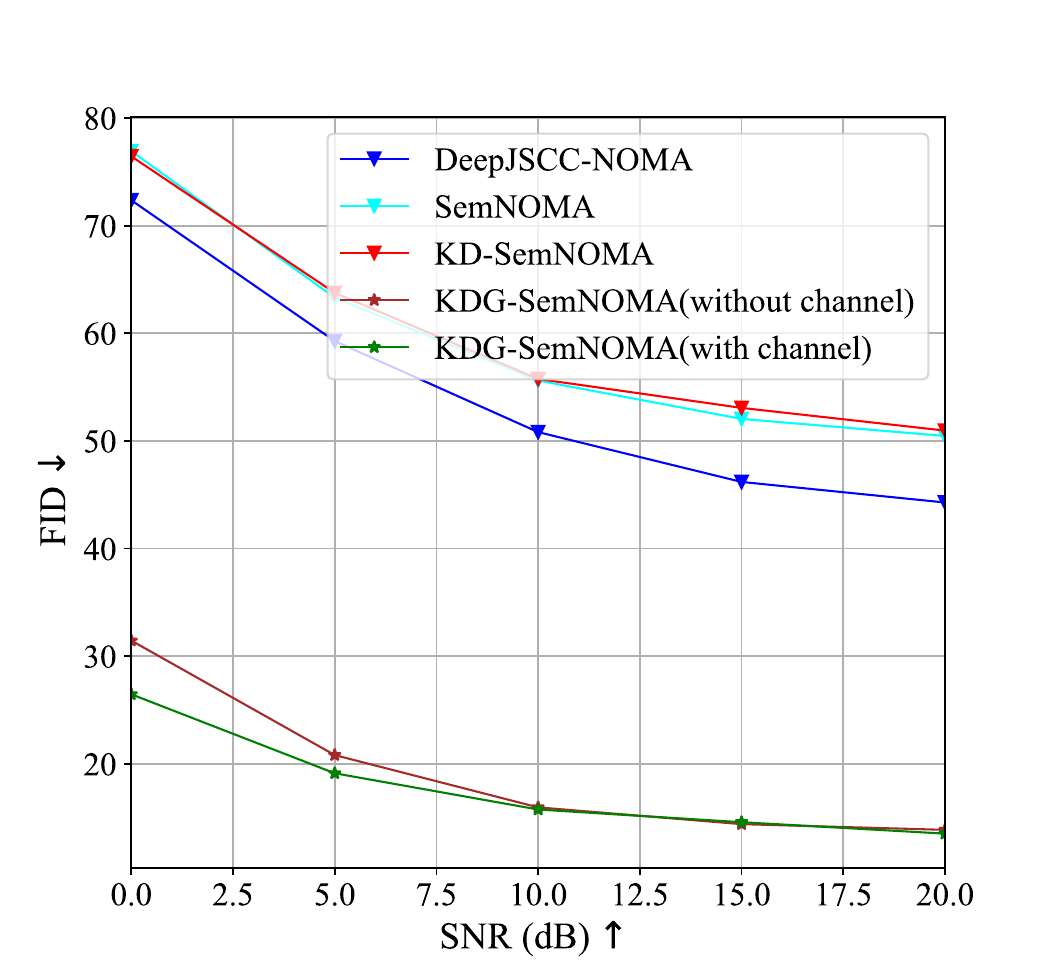}
			\captionsetup{font={footnotesize}}
			\caption{FID vs. SNR (AWGN)}
			\label{fig8b}
		\end{subfigure}
		\hfill
		\begin{subfigure}[t]{0.48\columnwidth}
			\centering
			\includegraphics[width=\linewidth]{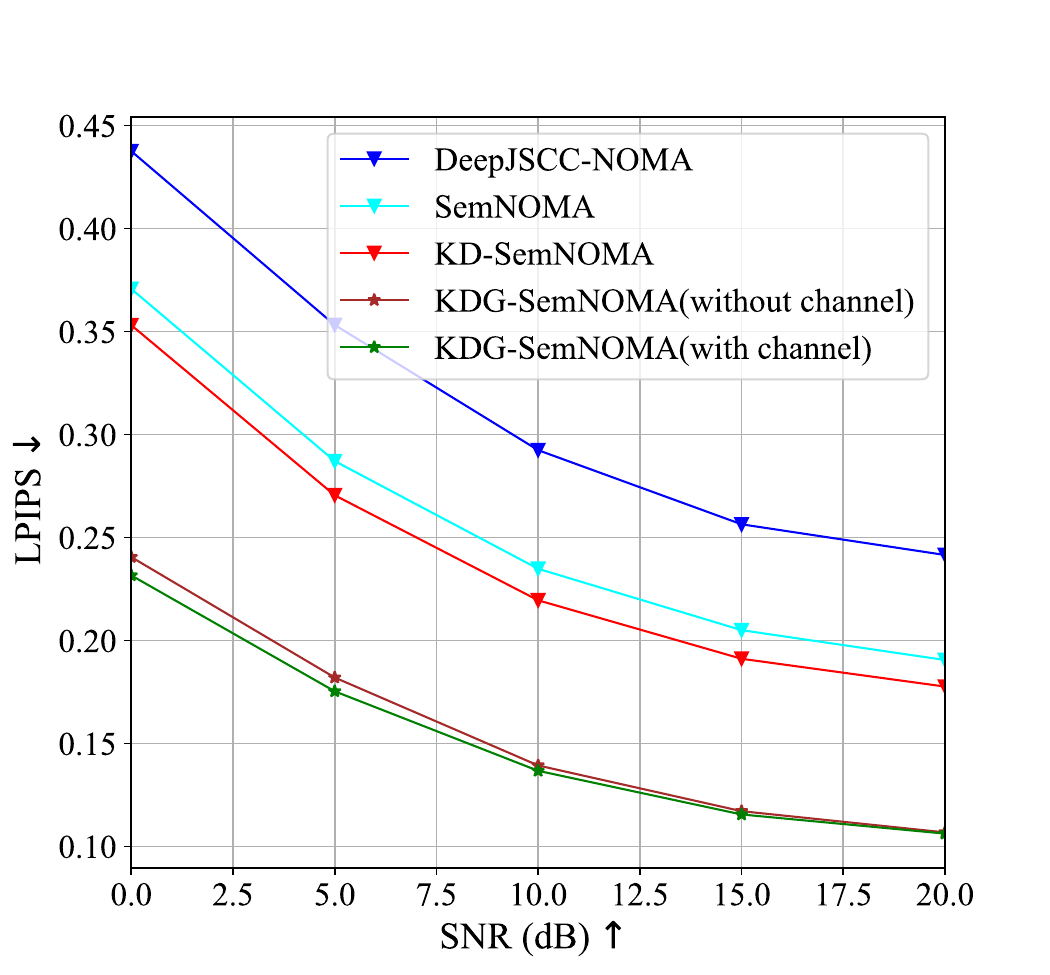}
			\captionsetup{font={footnotesize}}
			\caption{LPIPS vs. SNR (Rayleigh)}
			\label{fig7a}
		\end{subfigure}
		\hfill
		\begin{subfigure}[t]{0.48\columnwidth}
			\centering
			\includegraphics[width=\linewidth]{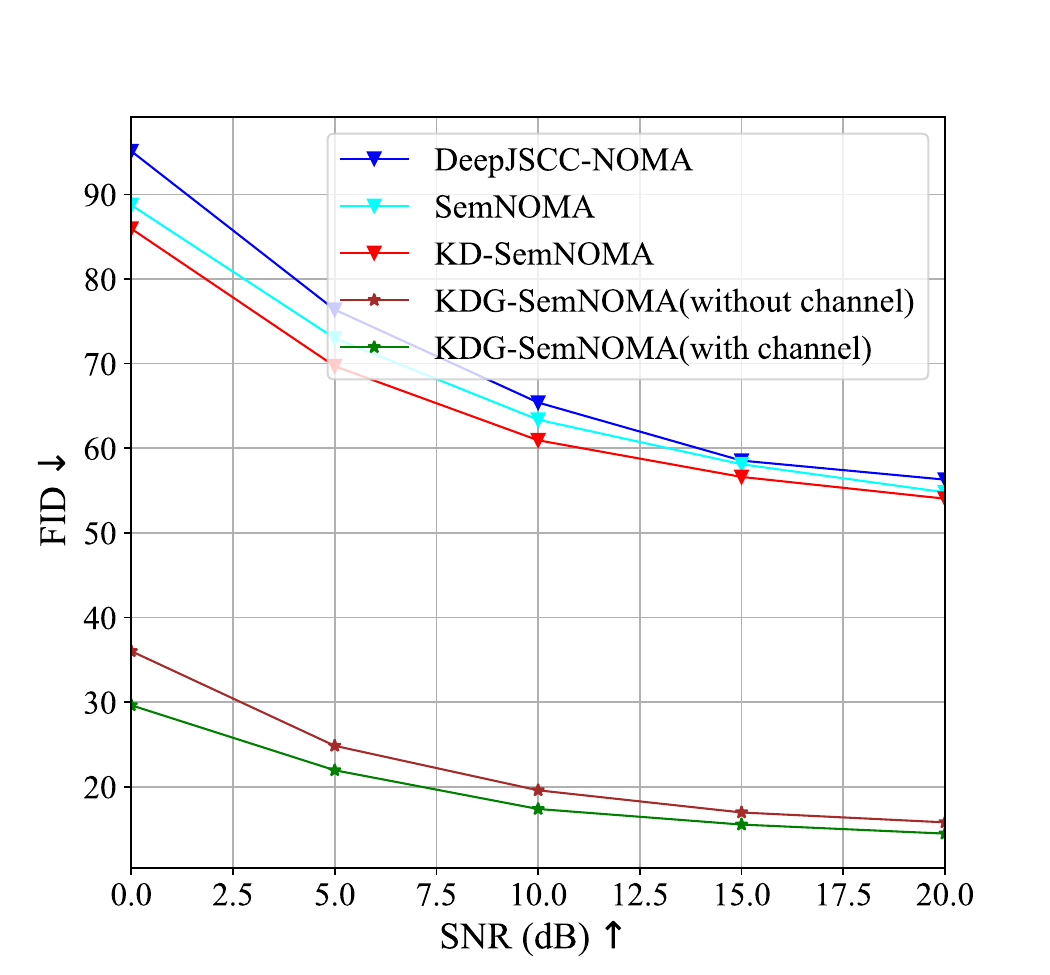}
			\captionsetup{font={footnotesize}}
			\caption{FID vs. SNR (Rayleigh)}
			\label{fig7b}
		\end{subfigure}
		\captionsetup{font={footnotesize,{color=black}}}
		\caption{Perceptual quality comparison under AWGN and Rayleigh fading channels (2UE, $\rho=1/48$). Subfigures (a) and (b) show LPIPS and FID versus SNR under AWGN channel, respectively, while subfigures (c) and (d) show LPIPS and FID under Rayleigh fading channel.}
		\label{fig:performance_lpips_fid_ffhq_M32}
		\vspace{-6.8mm}
	\end{figure*}
	\subsubsection{Performance analysis of KD-SemNOMA}
	We evaluate the reconstruction fidelity (PSNR) on the high-resolution FFHQ-256 dataset with a compression ratio of $\rho = 1/48$. The simulation results under AWGN and Rayleigh fading channels are illustrated in Fig.~\ref{fig:performance_psnr}. The proposed ConvNeXt-based \textbf{SemNOMA} framework outperform the ResNet-based \textbf{DeepJSCC-NOMA} by $0.3 \sim 0.4$ dB. The proposed knowledge distillation strategy further boosts performance, yielding an additional gain of $0.3 \sim 0.5$ dB in \textbf{KD-SemNOMA}. Notably, at the same transmission overhead, our non-orthogonal \textbf{KD-SemNOMA} scheme outperforms the orthogonal \textbf{SemOMA} benchmark by approximately $0.3 \sim 0.5$ dB, verifying the efficacy of the proposed teacher-student optimization in mitigating NOMA interference. In the more challenging Rayleigh fading scenarios, the advantages of our framework are even more pronounced. Thanks to the channel-aware Enhanced AF-Module, \textbf{SemNOMA} achieves a substantial gain of $1 \sim 1.5$ dB over the baseline \textbf{DeepJSCC-NOMA}. With KD optimization, \textbf{KD-SemNOMA} obtains a further $0.4$ dB improvement.
	Compared to the interference-free \textbf{SemOMA} (same overhead), \textbf{KD-SemNOMA} shows competitive performance: while slightly lower in PSNR in the low-SNR regime ($[0, 6]$ dB), it surpasses \textbf{SemOMA} in the high-SNR regime ($[6, 20]$ dB). A critical advantage of the proposed semantic communication framework is its graceful performance degradation. As observed in Fig.~\ref{fig:performance_psnr}, the traditional separation-based scheme \textbf{BPG+LDPC+QAM+SIC} exhibits a severe ``cliff effect.'' Its reconstruction quality drops sharply when the channel quality falls below a threshold (approx. 5-6 dB for AWGN and 10-11 dB for Rayleigh), leading to complete decoding failure. In contrast, our proposed schemes maintain intelligible image reconstruction even in extremely low-SNR regimes, demonstrating superior robustness.

	\subsubsection{Effectiveness of Stage-II GAN Refinement}
	To evaluate the visual realism of the reconstructed images, we analyze the LPIPS and FID metrics under both AWGN and Rayleigh fading channels, as illustrated in Fig.~\ref{fig:performance_lpips_fid_ffhq_M32}.
	It is observed that the one-stage methods (\textbf{DeepJSCC-NOMA}, \textbf{SemNOMA}, and \textbf{KD-SemNOMA}), which are optimized solely based on pixel-wise MAE loss, exhibit relatively high LPIPS and FID scores. This corroborates the well-known issue where pixel-level objectives tend to produce over-smoothed images with missing high-frequency textures.
	In contrast, the proposed two-stage \textbf{KDG-SemNOMA} framework significantly reduces both LPIPS and FID values across the entire SNR range. By incorporating the adversarial loss and perceptual loss in the second stage, the cGAN effectively hallucinates realistic details and restores facial textures that were lost during the semantic compression and transmission, thereby aligning the reconstruction quality more closely with human perception.

	\subsubsection{Impact of Channel Conditioning}
	To validate the necessity of the channel-conditional design, we compare the proposed scheme with ablation variant, \textbf{KDG-SemNOMA (w/o channel)}, where the generator receives only image features without explicit CSI. As observed in Fig.~\ref{fig:performance_lpips_fid_ffhq_M32}, the conditional model consistently outperforms the unconditional variant. This performance gap is particularly pronounced in low-SNR regimes and complex Rayleigh fading channels, indicating that without channel guidance, the network struggles to distinguish between severe noise and intrinsic textures. Furthermore, explicit channel conditioning acts as a strong structural prior, which improves robustness against complex channel environments, and also leads to more stable training convergence compared to the unconditional counterpart.
	\vspace{-3.5mm}
	\section{Acknowledgement}
	The work was supported in part by the Natural Science Foundation of China (NSFC) under Grant 62471036, in part by Shandong Province Natural Science Foundation under Grant ZR2025QA30, in part by Beijing Natural Science Foundation under Grants L242011, QY24167, QY25256, QY25257.

	\vspace{-3.0mm}
	\section{Conclusion} \label{conclusion}
	This paper proposed KDG-SemNOMA, a robust multi-user semantic communication framework tailored for 6G-empowered RV networks. By integrating channel-adaptive coding, knowledge distillation, and conditional GAN refinement, our approach effectively mitigates NOMA interference and enhances visual perceptual quality. Simulation results on FFHQ-256 demonstrate significant gains over state-of-the-art baselines in both pixel-level fidelity and visual realism, offering a promising solution for 6G sustainable and intelligent autonomous mobility systems.

	\vspace{-3mm}

\end{document}